\documentclass[prd,a4paper,showpacs]{revtex4}
\usepackage{epsfig}

\usepackage{amssymb}
\usepackage{amsfonts}
\usepackage{graphicx}
\usepackage{keyval,graphicx}
\usepackage{textcomp,wasysym}

\begin{document}

\bibliographystyle{unsrt}


\title{The evasion of helicity selection rule in $\chi_{c1}\to VV$ and $\chi_{c2}\to VP$ via intermediate charmed meson loops}

\author{Xiao-Hai Liu$^1$\footnote{xhliu@ihep.ac.cn}, Qiang Zhao$^{1,2}$\footnote{zhaoq@ihep.ac.cn}}

\affiliation{1) Institute of High Energy Physics, Chinese Academy of
Sciences, Beijing 100049, P.R. China \\
2) Theoretical Physics Center for Science Facilities, CAS, Beijing
100049, China}

\date{\today}

\begin{abstract}
The hadronic decays of  $\chi_{c1}\to VV$ and $\chi_{c2}\to VP$ are
supposed to be suppressed by the helicity selection rule in the pQCD
framework. With an effective Lagrangian method, we show that the
intermediate charmed meson loops can provide a mechanism for the
evasion of the helicity selection rule, and result in sizeable decay
branching ratios in some of those channels. The theoretical
predictions can be examined by the forthcoming BES-III data in the
near future.

\end{abstract}

\maketitle

\section{Introduction}

The exclusive decays of heavy quarkonium have been an important
platform for studying the nature of strong interactions in the
literature~\cite{Brodsky:1981kj,Chernyak:1981zz,Chernyak:1983ej,Brambilla:2004wf,Voloshin:2007dx,bes-iii}
since the discovery of quantum chromodynamics (QCD). In this energy
region relatively large energy scale ($\thicksim m_c, m_b$) is
involved such that some perturbative QCD(pQCD)  asymptotic behaviors
can be expected, for instance, the so-called \textit{helicity
selection rule} \cite{Chernyak:1981zz}. By studying the
manifestation of the helicity selection rule in heavy quarkonium
decays, we expect to gain insights into the property of QCD in the
interplay of the perturbative and non-perturbative regime.

We briefly review this powerful method that is elaborated on in Ref.
\cite{Chernyak:1981zz}. For a charmonium meson $J_{c\bar{c}}$
decaying into two light mesons $h_1$ and $h_2$, the perturbative
method gives the asymptotic behavior of the branching ratio as
follows
\begin{eqnarray}
BR_{ J_{c\bar{c}} (\lambda)\to h_1(\lambda_1) h_2(\lambda_2) } \sim
\left(\frac{\Lambda_{QCD}^2}{m_c^2} \right)^{|\lambda_1+\lambda_2|
+2},
\end{eqnarray}
where $\lambda$, $\lambda_1$, and $\lambda_2$ are the helicities of
the corresponding mesons. This is the result of the pQCD method to
leading-twist accuracy; i.e. only the valence Fock state (here it is
$c\bar{c}$) is considered. It is obvious that the leading
contribution comes from when $\lambda_1+\lambda_2=0$, (while the
higher twist will be suppressed by at least a factor of
$\Lambda_{QCD}^2/ m_c^2$.) while the helicity configurations which
do not satisfy this relation will be suppressed.

An alternative description of this selection rule is to depict it
with the quantum number ``naturalness", which is defined as $\sigma
\equiv P(-1)^J$, where $P$ and $J$ are the parity and spin of the
particle, respectively. The selection rule then requires that
\begin{equation}
 \sigma ^{initial}= \sigma_1 \sigma_2 \ .
\end{equation}
That is, the initial-state naturalness should be equal to the
product of those of the final states. We can comprehend this in such
a way that, if $\sigma ^{initial}\neq \sigma_1 \sigma_2$, to keep
the parity conservation and Lorentz invariance, the amplitude will
contain a Levi-Civita tensor $\epsilon_{\mu\nu\alpha\beta}$, which
will be contracted with the polarization vectors and momenta of the
involved mesons. For instance, for the process $\chi_{c1}\to VV$
where $VV$ represent a pair of vector mesons, in the rest frame of
$\chi_{c1}$, the non-vanishing covariant amplitude will require one
vector meson be transversely polarized and the other one is
longitudinally polarized. This leads to $\lambda_1+\lambda_2\neq 0$,
which will violate the helicity selection rule and is supposed to be
suppressed. Another well-known example is the process $J/\psi\to
VP$, where the non-vanishing covariant amplitude also violates the
selection rule.

The helicity selection rule has been used for studying some
exclusive decay processes of heavy quarkonia. In
Refs.~\cite{Chernyak:1981zz,Chernyak:1983ej,Feldmann:2000hs}, some
allowed or suppressed processes are explicitly listed.
Interestingly, with the accumulation of the experimental data, more
and more observations suggest significant discrepancies between the
data and the selection-rule expectations. For instance, the decays
of $J/\psi \to VP$ and $\eta_c \to VV$ would be suppressed by this
rule. In reality, they are rather important decay channels for
$J/\psi$ and $\eta_c$, respectively~\cite{Amsler:2008zzb}. One
possible reason why the perturbative method fails here could be that
although the mass of the charm quark is heavy, it is, however, not
as heavy as pQCD demands. Therefore, it is not safe to apply the
helicity selection rule to charmonium decays, while the situation
may be improved in bottomonium decays. Taking into account these
issues, it implies that there might be some other mechanisms that
can contribute to those helicity-selection-rule-forbidden processes,
such as higher order contributions, final state interactions, or
some other long-distance effects. Theoretical discussions on these
mechanisms in $B$-meson or heavy quarkonium decays can be found in
the
literature~\cite{Colangelo:1989gi,Ciuchini:1997hb,Isola:2001ar,Isola:2001bn,Cheng:2004ru,
Colangelo:2003sa,Zhao:2008eg,Zhang:2009kr,Liu:2009iw,Santorelli:2007xg,Gong:2008ue}.

In Ref.~\cite{Zhao:2008eg,Zhang:2009kr}, the role played by the
intermediate charmed meson loops in the understanding of the
so-called ``$\rho$-$\pi$ puzzle" and $\psi(3770)$ non-$D\bar{D}$
decays into $VP$ was studied. It shows that the long-distance
interactions via the intermediate meson loops provide a possible
mechanism for understanding the deviations from pQCD expectations.
In the present work, we will study the $P$-wave charmonium decays,
i.e. $\chi_{c1}\to VV$ and $\chi_{c2}\to VP$, which are suppressed
by the helicity selection rule but may be enhanced by the
intermediate charmed meson loop transitions. The branching ratio of
$\chi_{c1}\to K^{*0}\bar{K}^{*0}$  at order of $10^{-3}$ has been
measured by experiment~\cite{Amsler:2008zzb}. According to the SU(3)
flavor symmetry, it is justified to predict some other channels as
$\rho\rho$, $\omega\omega$ and $\phi\phi$, of which the branching
ratios may also be sizeable. Therefore, it is interesting to
investigate the intermediate meson loop transitions as a mechanism
for evading the helicity selection rule. We also note that the
process $\chi_{c2}\to VP$ will be further suppressed by the
approximate $G$-parity or isospin ($U$-spin for strange mesons)
conservation, and the $\chi_{c2}$ decays into neutral $VP$ with
fixed $C$ parities are forbidden by C-parity conservation. By
comparing the results with the SU(3) flavor symmetry expectation, we
anticipate that the non-perturbative mechanism can be highlighted.

It is worth noting that for the P-wave charmonium exclusive decays,
the next higher Fock state $c\bar{c}g$, i.e. the so-called
color-octet state, will contribute at the same order as the
color-singlet state $c\bar{c}$ in the framework of perturbative
method
\cite{Feldmann:2000hs,Cho:1995vh,Bolz:1997ez,Bodwin:1992ye,Bodwin:1994jh}.
This scenario may share the same intrinsic physics with the
intermediate meson loop transitions, but based on different view
angles. Namely, one description is at the quark-gluon level, while
the other is at the hadron level. Taking into account the relatively
low energy scale, it is not easy to handle these processes with the
factorization method for the quark-gluon interactions. Some
qualitative discussions can be found in
\cite{Isola:2001ar,Cheng:2004ru,Colangelo:2003sa} and references
therein.

This paper is arranged as follows: In Sec. II, the formulae for the
intermediate meson loops as a long-distance effect in the charmonium
decays will be given by an effective Lagrangian method. The
numerical results and discussions will be given in Sec. III. The
conclusion will be drawn in Sec. IV.

\section{Long-distance contribution via intermediate meson loops}

In Figs.~\ref{chic1},~\ref{chic2rho} and~\ref{chic2k}, the
intermediate meson loops are plotted for $\chi_{c1}\to VV$,
$\chi_{c2}\to \rho\pi$ and $\chi_{c2}\to K^*\bar{K}$, respectively.
%
The relevant effective Lagrangians that will be used are based on
the heavy quark symmetry and $SU(3)$ symmetry
\cite{Cheng:2004ru,Colangelo:2003sa,Casalbuoni:1996pg}. The spin
multiplet for these four $P$-wave charmonium states are expressed as
\begin{eqnarray}
P_{c\bar{c}}^\mu = \left( \frac{1+ \rlap{/}{v} }{2} \right)
\left(\chi_{c2}^{\mu\alpha}\gamma_{\alpha} +\frac{1}{\sqrt{2}}
\epsilon_{\mu\nu\alpha\beta}v^{\alpha}\gamma^{\beta}\chi_{c1}^\nu
+\frac{1}{\sqrt{3}}(\gamma^\mu -v^\mu)\chi_{c0} +h_c^\mu \gamma_{5}
\right)
\left( \frac{1- \rlap{/}{v} }{2} \right).
\end{eqnarray}
The charmed and anti-charmed meson triplet read
\begin{eqnarray}
H_{1i}&=&\left( \frac{1+ \rlap{/}{v} }{2} \right)
[\mathcal{D}_i^{*\mu}
\gamma_\mu -\mathcal{D}_i\gamma_5], \\
H_{2i}&=& [\bar{\mathcal{D}}_i^{*\mu} \gamma_\mu
-\bar{\mathcal{D}}_i\gamma_5]\left( \frac{1- \rlap{/}{v} }{2}
\right),
\end{eqnarray}
where $\mathcal{D}$ and $\mathcal{D}^*$ denote the pseudoscalar and
vector charmed meson fields respectively, i.e.
$\mathcal{D}^{(*)}=(D^{0(*)},D^{+(*)},D_s^{+(*)})$. Then the
effective Lagrangian that describes the interaction between the
P-wave charmonium and the charmed mesons reads
\begin{eqnarray}
\mathcal{L}_1=i g_1 Tr[P_{c\bar{c}}^\mu \bar{H}_{2i}\gamma_\mu
\bar{H}_{1i}] + H.c.
\end{eqnarray}
The effective Lagrangians describe the couplings of charmed mesons
to light hadrons read
\begin{eqnarray}
\mathcal{L}_{\mathcal{D}\mathcal{D}\mathcal{V}}&=& -i
g_{DDV}\bar{\mathcal{D}}_i
{\stackrel{\leftrightarrow}{\partial}}_\mu \mathcal{D}_j
(\mathcal{V}^\mu)_{ij}, \\
\mathcal{L}_{\mathcal{D}^*\mathcal{D}\mathcal{V}} &=& -2f_{D^*DV}
\epsilon_{\mu\nu\alpha\beta}(\partial^\mu \mathcal{V}^\nu)_{ij}
(\bar{\mathcal{D}}_i {\stackrel{\leftrightarrow}{\partial}^\alpha}
\mathcal{D}^{*\beta}_j - \bar{\mathcal{D}}^{*\beta}_i
{\stackrel{\leftrightarrow}{\partial}^\alpha} \mathcal{D}_j ) , \\
\mathcal{L}_{\mathcal{D}^*\mathcal{D}^*\mathcal{V}} &=& i
g_{D^*D^*V} \bar{\mathcal{D}}^{*\nu}_i
{\stackrel{\leftrightarrow}{\partial}_\mu} \mathcal{D}^*_{j\nu}
(\mathcal{V}^\mu)_{ij} + 4i f_{D^*D^*V} \bar{\mathcal{D}}^{*\mu}_i
(\partial_\mu \mathcal{V}_\nu - \partial_\nu \mathcal{V}_\mu)_{ij}
\mathcal{D}^{*\nu}_j, \\
\mathcal{L}_{\mathcal{D}^*\mathcal{D}\mathcal{P}} &=& -i g_{D^*DP}
(\bar{\mathcal{D}}_i \partial_\mu \mathcal{P}_{ij}
\mathcal{D}^{*\mu}_j - \bar{\mathcal{D}}^{*\mu}_i \partial_\mu
\mathcal{P}_{ij} \mathcal{D}_j), \\
\mathcal{L}_{\mathcal{D}^*\mathcal{D}^*\mathcal{P}} &=& \frac{1}{2}
g_{D^*D^*P}\epsilon_{\mu\nu\alpha\beta} \bar{\mathcal{D}}^{*\mu}_i
\partial^\nu \mathcal{P}_{ij} {\stackrel{\leftrightarrow}{\partial}^\alpha}
\mathcal{D}^{*\beta}_j,
\end{eqnarray}
where $\mathcal{V}$ and $\mathcal{P}$ are the pseudoscalar octet and
vector nonet, and we take the following conventions
\begin{eqnarray}
\mathcal{V}= \left(
  \begin{array}{ccc}
    \frac{1}{\sqrt{2}}(\rho^0+\omega) & \rho^+ & K^{*+} \\
    \rho^- & \frac{1}{\sqrt{2}}(-\rho^0+\omega)  & K^{*0} \\
    K^{*-} & \bar{K}^{*0} & \phi \\
  \end{array}
\right),
\end{eqnarray}
\begin{eqnarray}
\mathcal{P}= \left(
  \begin{array}{ccc}
    \frac{1}{\sqrt{2}}(\pi^0+\eta) & \pi^+ & K^{+} \\
    \pi^- & \frac{1}{\sqrt{2}}(-\pi^0+\eta)  & K^{0} \\
    K^{-} & \bar{K}^{0} & -\sqrt{\frac{2}{3}}\eta \\
  \end{array}
\right).
\end{eqnarray}

\subsection{$\chi_{c1}\to VV$}

\begin{figure}[t]
\includegraphics[width=0.5\hsize]{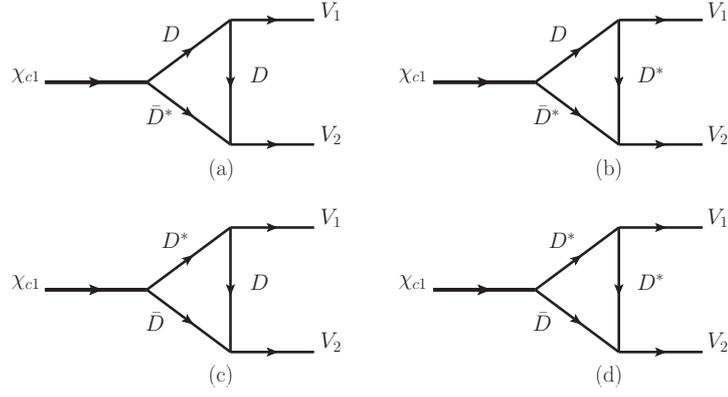}
\caption{Triangle loop diagrams that describe the long-distance
contributions in $\chi_{c1}\to V_1 V_2$, where $V_1 V_2$ represent
$K^*\bar{K}^*$, $\rho\rho$, $\omega\omega$, and $\phi\phi$,
respectively. }\label{chic1}
\end{figure}

Figure~\ref{chic1} illustrates the long-distance contributions for
$\chi_{c1}\to VV$. One notices that there is no such a vertex of
$\chi_{c1}D^*\bar{D}^*$ in Fig.~\ref{chic1}. It is because to
conserve parity and keep Lorentz invariance, the relative angular
momentum between $D^*\bar{D}^*$ should be $L=2$, and the total spin
$S=2$. But such kind of coupling is not present in the expansion of
the effective Lagrangian $\mathcal{L}_1$. For Fig.~\ref{chic1}(a)
and (b), we take the convention of the momenta as $\chi_{c1}(p)\to
D(q_1)\bar{D}^*(q_2)[D^{(*)}(q)]\to V_1(p_1) V_2(p_2)$, where the
exchanged particle between $D$ and $\bar{D}^*$ is indicated in the
square bracket. Figure~\ref{chic1}(c) and (d) will give the same
contribution as Fig.~\ref{chic1}(a) and (b). For simplicity, we just
write down the amplitude $\mathcal{M}_{1a}$ and $\mathcal{M}_{1b}$
and take the final states $\rho^+\rho^-$ as an example:
\begin{eqnarray}
\mathcal{M}_{1a} &=& 2ig_{DD^*\chi_{c1}} g_{DDV} f_{D^*DV}
\epsilon^{\chi_{c1}}_\lambda \epsilon_1^{*\sigma} \epsilon_2^{*\tau}
\int
\frac{d^4 q}{(2\pi)^4}  \nonumber \\
&\times& (q_{1\sigma}+q_{\sigma}) \epsilon_{\mu\tau\alpha\beta}
p_2^\mu (q^\alpha-q_2^\alpha)  \frac{g^{\lambda\beta} -
q_{2}^\lambda q_{2}^\beta/m_{D^*}^2}{D_a D_1 D_2} \mathcal{F}(q^2),
\\
\mathcal{M}_{1b} &=& 2ig_{DD^*\chi_{c1}} g_{DDV} f_{D^*DV}
\epsilon^{\chi_{c1}}_\lambda \epsilon_1^{*\sigma} \epsilon_2^{*\tau}
\int
\frac{d^4 q}{(2\pi)^4}  \nonumber \\
&\times& \epsilon_{\mu\sigma\alpha\beta} p_1^\mu(q_1^\alpha
+q^\alpha) \left[ g_{D^*D^*V}(q_{2\tau} -q_\tau) g_{\gamma\delta}+ 4
f_{D^*D^*V}(p_{2\delta} g_{\tau\gamma}-p_{2\gamma} g_{\delta\tau})
\right] \nonumber \\
&\times& (g^{\beta\gamma}-q^\beta
q^\gamma/m_{D^*}^2)(g^{\lambda\delta}-q_2^\lambda
q_2^\delta/m_{D^*}^2) \times \frac{1}{D_b D_1 D_2} \mathcal{F}(q^2),
\end{eqnarray}
where $D_a=q^2-m_D^2$, $D_b=q^2-m_{D^*}^2$, $D_1=q_1^2-m_D^2$, and
$D_2=q_2^2-m_{D^*}^2$. The results of other channels can be obtained
similarly.

Since the mass of $\chi_{c1}$ is under the threshold of
$D\bar{D}^*$, intermediate mesons $D$ and $\bar{D}^*$ can not be
on-shell simultaneously. We phenomenally introduce a form factor
$\mathcal{F}(q^2)$ as  has been done in Refs.
\cite{Cheng:2004ru,Colangelo:2003sa},
\begin{eqnarray}
\mathcal{F}(q^2)=\prod\limits_i
\left(\frac{m_i^2-\Lambda_i^2}{q_i^2-\Lambda_i^2} \right),
\end{eqnarray}
where $q_i=q,\ q_1,\ q_2$. The cut-off energy is chosen as
$\Lambda_i=m_i +\alpha \Lambda_{QCD}$, and $m_i$ is the mass of the
corresponding exchanged particle. This is somehow different from the
form factor adopted in Refs. \cite{Cheng:2004ru,Colangelo:2003sa}.
We should mention that the form factor is necessary for killing the
divergence of the loop integrals, although it will also give rise to
the model-dependent aspects of the calculations. We will discuss
this in detail later.

\subsection{$\chi_{c2}\to VP$}

\begin{figure}[t]
\includegraphics[width=0.5\hsize]{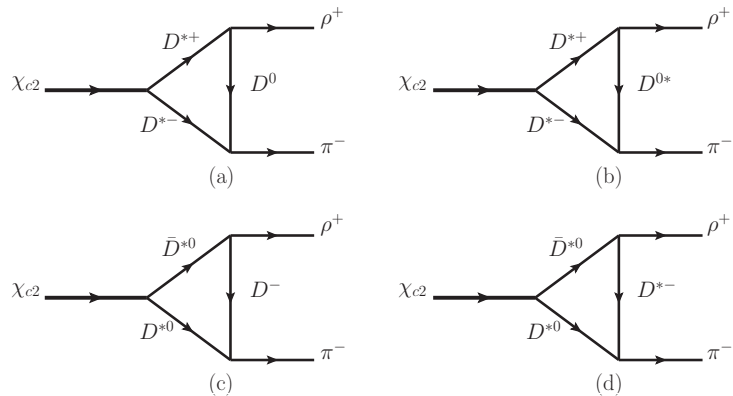}
\caption{Triangle loop diagrams that describe the long-distance
contributions in $\chi_{c2}\to \rho^+\pi^-$. The diagrams for
$\chi_{c2}\to \rho^-\pi^+$ are implicated. }\label{chic2rho}
\end{figure}

\begin{figure}[t]
\includegraphics[width=0.5\hsize]{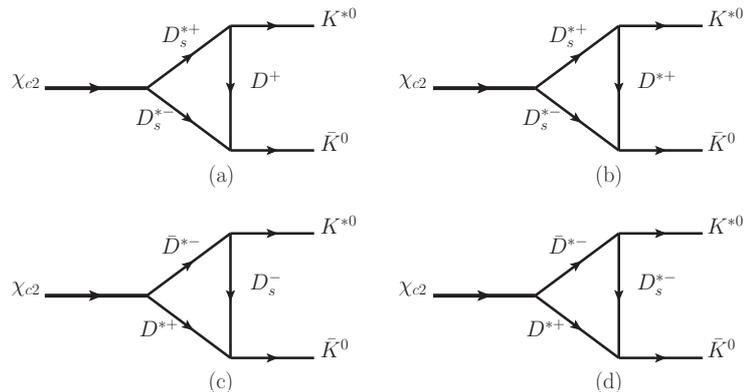}
\caption{Triangle loop diagrams that describe the long-distance
contributions in $\chi_{c2}\to K^{*0}\bar{K}^0$. The diagrams for
$\chi_{c2}\to \bar{K}^{*0} {K}^0$ are implicated. }\label{chic2k}
\end{figure}

As mentioned previously, the decay $\chi_{c2}\to VP$ suffers not
only from the suppression of the helicity selection rule, but also
from the approximate $G$-parity or isospin/$U$-spin conservation.
However, because of the relatively large mass difference between the
$u$/$d$ quark and $s$ quark, the intermediate meson loops may still
bring in sizeable branching ratios for $\chi_{c2}\to K\bar{K}^*
+c.c.$. Next, we will consider, $\chi_{c2}\to \rho^+\pi^- +c.c.$ and
$\chi_{c2}\to K^*\bar{K} +c.c.$. The $\chi_{c2}$ decays into neutral
$VP$ with fixed $C$-parity, such as $\rho^0\pi^0$ and $\omega\eta$
etc.,  forbidden by $C$-parity conservation. The loop diagrams for
these two processes are presented in Fig.~\ref{chic2rho} and
Fig.~\ref{chic2k}, respectively. The convention of the momenta
follows that of Fig.~\ref{chic1}.

The relative signs between the following amplitudes are opposite,
i.e. $\mathcal{M}_{2a}$ and $\mathcal{M}_{2c}$, $\mathcal{M}_{2b}$
and $\mathcal{M}_{2d}$, $\mathcal{M}_{3a}$ and $\mathcal{M}_{3c}$,
and $\mathcal{M}_{3b}$ and $\mathcal{M}_{3d}$. This leads to
destructive interferences between these amplitudes, and is a
reflection of the approximate isospin or $U$-spin invariance. We
explicitly list the amplitudes, $\mathcal{M}_{2a}$ and
$\mathcal{M}_{2b}$,
\begin{eqnarray}
\mathcal{M}_{2a} &=& 2i g_{D^*D^*\chi_{c2}} f_{D^*DV} g_{D^*DP}
\epsilon_{\xi\eta}^{\chi_{c2}} \epsilon_{\rho^+}^\nu \int \frac{d^4
q}{(2\pi)^4}  \epsilon_{\mu\nu\alpha\beta} p_1^\mu(q_1^\alpha
+q^\alpha)
p_2^\lambda \nonumber \\
&\times& (g^{\xi\beta}-q_1^\xi
q_1^\beta/m_{D^*}^2)(g^{\eta\lambda}-q_2^\eta q_2^\lambda/m_{D^*}^2)
 \frac{1}{D_a D_1 D_2} \mathcal{F}(q^2), \\
\mathcal{M}_{2b} &=& -\frac{1}{2}i g_{D^*D^*\chi_{c2}} g_{D^*D^*P}
\epsilon_{\xi\eta}^{\chi_{c2}} \epsilon_{\rho^+}^\tau \int \frac{d^4
q}{(2\pi)^4} \epsilon_{\rho\sigma\alpha\beta}
p_2^\sigma(q^\alpha-q_2^\alpha) \nonumber \\
&\times& \left[-g_{D^*D^*V}(q_{1\tau}+q_\tau)g^{\gamma\delta}
-4f_{D^*D^*V}(p_1^\gamma g_\tau^\delta -p_1^\delta g^\gamma_\tau)
\right] \nonumber \\
&\times& (g^{\xi\gamma}-q_1^\xi q_1^\gamma/m_{D^*}^2)
(g^{\eta\beta}-q_2^\eta q_2^\beta/m_{D^*}^2)
(g^{\delta\rho}-q^\delta q^\rho/m_{D^*}^2)
 \frac{1}{D_b D_1 D_2}\mathcal{F}(q^2) \ ,
\end{eqnarray}
while the others can be obtained similarly. As for the polarization
sums of $\epsilon_{\xi\eta}^{\chi_{c2}}$, we follow the expressions
in Ref.~\cite{Zou:2002ar}.

\section{Numerical results}

Before proceeding to the numerical results, we first discuss the
parameters, such as the coupling constants, in the formulation. In
the chiral and heavy quark limit, the following relations can be
obtained~\cite{Cheng:2004ru,Casalbuoni:1996pg}:
\begin{eqnarray}
&&g_{DDV}=g_{D^*D^*V}=\frac{\beta g_V}{\sqrt{2}},\
f_{D^*DV}=\frac{f_{D^*D^*V}}{m_{D^*}}=\frac{\lambda g_V}{\sqrt{2}},
\nonumber \\
&&g_V=\frac{m_\rho}{f_\pi},\
g_{D^*D^*\pi}=\frac{g_{D^*D\pi}}{\sqrt{m_D
m_{D^*}}}=\frac{2}{f_\pi}g, \nonumber \\
&&g_{D^*D_s K}=\sqrt{\frac{m_{D_s}}{m_D}}g_{D^*D\pi},\ g_{D^*_s D
K}= \sqrt{\frac{m_{D_s^*}}{m_{D^*}}},
\end{eqnarray}
where $\beta$ and $\lambda$ are commonly taken as $\beta=0.9$,
$\lambda=0.56\ \mbox{GeV}^{-1}$
\cite{Cheng:2004ru,Liu:2009iw,Isola:2003fh}, while $f_\pi$ is the
pion decay constant. With the measured branching ratio of $D^*\to
D\pi$ by CLEO-c, the coupling $g$ is determined as $g=0.59$
\cite{Anastassov:2001cw}. In the heavy quark limit, the expansion of
the effective Lagrangian $\mathcal{L}_1$ leads to relations for the
$P$-wave charmonium couplings to the charmed mesons as follows:
\begin{eqnarray}
g_{DD^*\chi_{c1}}&=& 2\sqrt{2}g_1 \sqrt{m_D m_{D^*} m_{\chi_{c1}}},
\nonumber \\
g_{D^*D^*\chi_{c2}} &=& 4 g_1 m_{D^*} \sqrt{m_{\chi_{c2}}},
\nonumber \\
g_1 &=& -\sqrt{\frac{m_{\chi_{c0}}}{3}} \frac{1}{f_{\chi_{c0}} },
\end{eqnarray}
where $g_1$ has been related to the $\chi_{c0}$ decay constant
$f_{\chi_{c0}}$. It can be approximately determined by the QCD sum
rule method, i.e. $f_{\chi_{c0}} \simeq 0.51\ \mbox{GeV}$
\cite{Colangelo:2002mj}.

The form factor parameter $\alpha$ generally cannot be determined
from the first principle. It is usually taken to be order of unity,
and depends on the particular process. In this work, since the
branching ratio of $\chi_{c1}\to K^{*0}\bar{K}^{*0}$ has been
measured in experiment~\cite{Amsler:2008zzb}, we will adopt the data
to constrain the form factor parameter.

The loop integrals are calculated with the software package
LoopTools \cite{Hahn:1998yk}. In Tables~\ref{table1} and
~\ref{table2}, we display the numerical results for the branching
ratios of $\chi_{c1}\to VV$ and $\chi_{c2}\to VP$ at $\alpha=0.3\sim
 0.33$, which correspond to the lower and upper bounds of
$BR(\chi_{c1}\to K^{*0}\bar{K}^{*0})=(1.6\pm 0.4)\times
10^{-3}$~\cite{Amsler:2008zzb}.

For $\chi_{c1}\to VV$, the branching ratio of the $\rho\rho$
channels is significant, while the $\omega\omega$ and $\phi\phi$
channel are relatively small. We also include the predictions from
the SU(3) flavor symmetry as a comparison. The parametrization is
given in the Appendix. It is interesting to see that the results
from the SU(3) flavor symmetry are basically compatible with those
given by the intermediate meson loops except that the branching
ratio for $\phi\phi$ is larger. It should be an indication for the
SU(3) flavor symmetry breaking. With the SU(3) symmetry breaking
parameter $R\simeq f_\pi/f_K = 0.838$~\cite{Amsler:2008zzb}, the
branching ratios agree with the intermediate meson loop results
pretty well.

Some insights into the transition mechanisms can be gained here:

i) The SU(3) flavor symmetry breaking $R = 0.838$ is consistent with
the loop transition behavior. Note that the intermediate $D_s$ and
$D_s^*$ pair has higher mass threshold, and the production of the
$\phi\phi$ will be relatively suppressed in comparison with the
non-strange $\rho\pi$ and $\omega\omega$ apart from the final-state
phase space differences. This corresponds to the flavor symmetry
breaking at leading order.

ii) The consistencies support the idea that the intermediate meson
loops provide a nature mechanism for the evasion of the helicity
selection rule as a long-distance transition. Since the
$\chi_{c0,1,2}$ are $P$-wave states, the short-distance transition
probes the first derivative of the $c\bar{c}$ wavefunction at the
origin, which however, would be suppressed in $\chi_{c1}\to VV$ and
$\chi_{c2}\to VP$ due to the helicity selection rule. By
annihilating the $c\bar{c}$ at long distance via the intermediate
meson loops, the helicity selection rule is then evaded.

iii) This phenomenon is slightly different from the $S$-wave
charmonium decays, such as $J/\psi\to \rho\pi$ etc, where it is not
easy to separate out the long-distance transitions from the
short-distance ones arising from a possible anomalous component of
wavefunction~\cite{Zhao:2008eg,Brodsky:1997fj,Rosner:2004wy}.

\begin{center}
\begin{table}
\begin{tabular}{|c|c|c|c|c|}
  \hline
  \hline
BR $(\times\mbox{10}^{-4})$ & $K^{*0}\bar{K}^{*0}$ & $\rho\rho$ & $\omega\omega$ & $\phi\phi$  \\
  \hline
  Exp. data & $16\pm 4$ & --- & --- & ---  \\
  \hline
  Meson loop & $12\sim 20$ & $26\sim 54$ & $8.7\sim 18$ & $2.7\sim 4.6$ \\
  \hline
  SU(3)$(R=1)$ & 16.0 & 26.8 & 8.8 & 6.8  \\
  \hline
  SU(3)$(R=0.838)$ & 16.0 & 32.0 & 10.6 & 4.0  \\
  \hline
\end{tabular}
\caption{Branching ratios for $\chi_{c1}\to VV$ predicted by the
intermediate meson loop transitions in the range $\alpha=0.3\sim
0.33$ corresponding to the measured lower and upper bound of
$BR(\chi_{c2}\to K^{*0}\bar{K}^{*0})$~\protect\cite{Amsler:2008zzb}.
Two results from the SU(3) flavor symmetry relation are presented
with $R=1$ and $R=0.838$. The long-dashed line means that
experimental data are unavailable.} \label{table1}
\end{table}
\end{center}

For $\chi_{c2}\to VP$, since there are no data available at this
moment, our predictions are based on the same form factor parameter
for $\chi_{c1}\to VV$. As listed in Tab.~\ref{table2}, the
accessible channels are only $K^*\bar{K} +c.c.$ and $\rho^+\pi^-
+c.c.$ Interestingly, the branching ratio for  $K^*\bar{K}+c.c.$
turns out to be sizeable, and the charged $\rho\pi$ is found much
smaller than the $K^*\bar{K}+c.c.$ channel. Qualitatively, this is
because that the cancellations between (a) and (c) (and similarly
between (b) and (d)) in Figs.~\ref{chic2rho} and \ref{chic2k} are
rather different. Namely, the $K^*\bar{K} +c.c.$ channel experiences
large $U$-spin symmetry breakings due to the significant mass
difference between $d$ and $s$ quark. In contrast, the $\rho\pi$
channel is originated from the $u$-$d$ mass difference. This unique
result can be examined by BES-III as a test of our model.

\begin{center}
\begin{table}
\begin{tabular}{|c|c|c|c|c|}
 \hline
  \hline
  BR$(\times 10^{-5})$  & $K^{*0}\bar{K}^0 +c.c.$ & $K^{*+}{K}^- +c.c.$ & $\rho^+\pi^- +c.c.$ \\
  \hline
  Meson loop & $4.0\sim 6.7$ & $4.0\sim 6.7$ & $(1.2\sim 2.0)\times 10^{-2}$ \\
  \hline
  Exp. data & --- & --- & --- \\
  \hline
\end{tabular}
\caption{Branching ratios for $\chi_{c2}\to VP$ predicted by the
intermediate meson loop transitions in the same range of
$\alpha=0.3\sim 0.33$. The long-dashed line means that experimental
data are unavailable. } \label{table2}
\end{table}
\end{center}

\begin{figure}[tb]
\includegraphics[width=0.5\hsize]{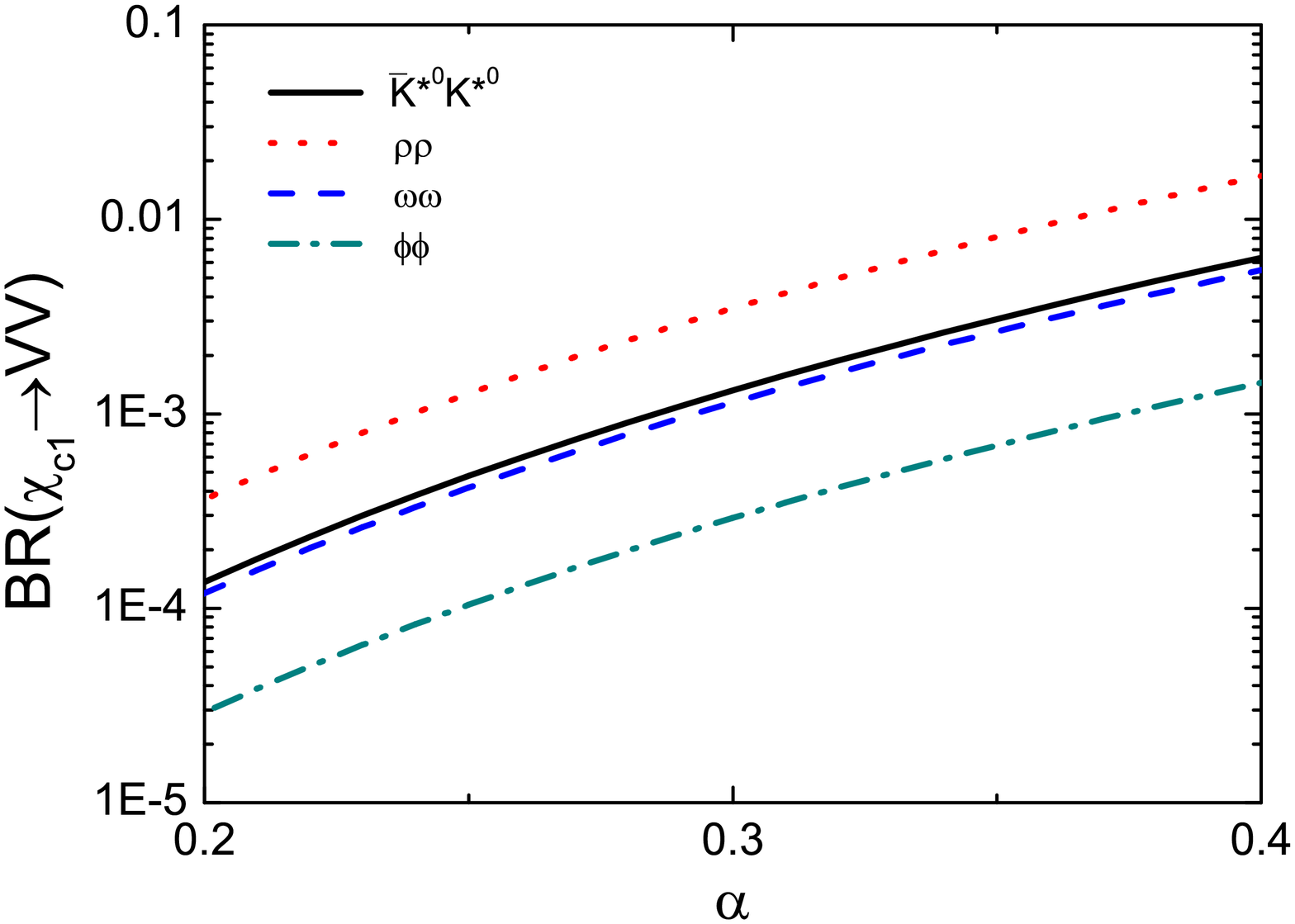}
\caption{ $\alpha$-dependence of the calculated branching ratios for
$\chi_{c1}\to VV$. }\label{brchic1}
\end{figure}

\begin{figure}[tb]
\includegraphics[width=0.5\hsize]{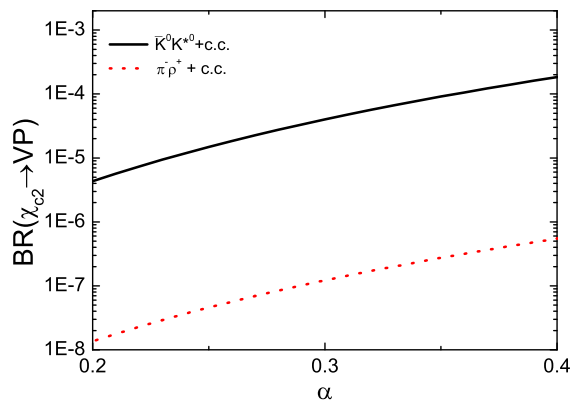}
\caption{  $\alpha$-dependence of the calculated branching ratios
for $\chi_{c2}\to VP$. }\label{brchic2}
\end{figure}

The sensitivities of the calculation results to the form factor
parameter $\alpha$ are presented in Figs.~\ref{brchic1}
and~\ref{brchic2}. One notices that the values for $\alpha$ in this
work are relatively smaller than those adopted in some other works
\cite{Cheng:2004ru,Liu:2009iw,Santorelli:2007xg}. This is acceptable
since in this work the form factor takes care of off-shell effects
arising from those three intermediate mesons, instead of only the
exchanged one in the
rescattering~\cite{Cheng:2004ru,Liu:2009iw,Santorelli:2007xg}. Given
the inevitable model-dependence introduced by the form factor, what
turns to be relatively stable and less model-dependent is the
relative branching ratio fractions among those decay channels within
the adopted range of $\alpha$. As a consequence, the theoretical
predictions for other decay channels can be better controlled by the
data for $\chi_{c1}\to K^{*0}\bar{K}^{*0}$.

\section{Conclusion}

In this work, we have discussed how the long-distance transitions
via the intermediate meson loops would contribute to the processes
$\chi_{c1}\to VV$ and $\chi_{c2}\to VP$, which are supposed to be
suppressed according to the helicity selection rule in QCD. With an
effective Lagrangian method with heavy quark and chiral symmetry,
this helicity-selection-rule evading mechanism is quantified.
Although there are still relatively large uncertainties arising from
the form factor parameter, we argue that the fewer sensitivities of
the branching ratio fractions among the accessible channels would
provide a better control of the theory predictions. Reasonable
ranges of the predictions are obtained.

We also compare the results from the intermediate meson loops with
the expectations of the SU(3) flavor symmetry, and find that they
are in a good agreement with each other. In particular, the SU(3)
flavor symmetry breaking would lead to a suppression of
$\chi_{c1}\to \phi\phi$ in comparison with the $\omega\omega$. It
can be well-understood by the heavier mass of the intermediate $D_s$
($D_s^*$) than the $D$ ($D^*$) state. Namely, the mass threshold of
the $D_s\bar{D_s^*}+c.c.$ is higher than $D\bar{D^*}+c.c.$ In
$\chi_{c2}\to VP$, the predicted branching ratio for the
$K^*\bar{K}+c.c.$ channel is at order of $10^{-5}$, while the
charged $\rho\pi$ channel is small. The suppression on the charged
$\rho\pi$ (in comparison with the $K^*\bar{K}+c.c.$) can be
comprehended as a consequence of the larger effects due to the
$U$-spin symmetry breaking rather than the isospin symmetry
breaking, i.e. $m_s-m_d >> m_d -m_u$.

In brief, we emphasize that the $P$-wave charmonium decay should be
ideal for examining the evading mechanisms of the helicity selection
rule. Our predictions can be examined by the high-statistics
$\chi_{cJ}$ production in the BES-III
experiment~\cite{bes-iii,maoyajun}.

\section*{Acknowledgement}

This work is supported, in part, by the National Natural Science
Foundation of China (Grants No. 10675131 and 10491306), Chinese
Academy of Sciences (KJCX3-SYW-N2), and Ministry of Science and
Technology of China (2009CB825200).

\section*{Appendix}

We adopt a simple
parametrization~\cite{Zhao:2005im,Zhao:2007ze,Li:2007ky} to study
$\chi_{c1}\to VV$ based on the SU(3) flavor symmetry. By assuming
that $\hat{H}$ represents the potential for the $c\bar{c}$
annihilating into gluons and then hadronizing into
$(q\bar{q})_{V_1}(q\bar{q})_{V_2}$, we define the transition
amplitude strength as
\begin{equation}
g_0\equiv \langle (q\bar{q})_{V_1}(q\bar{q})_{V_2} |\hat{H}
|\chi_{c1}\rangle \ ,
\end{equation}
where $q$ ($\bar{q}$) is a non-strange quark (anti-quark).
Considering the SU(3) flavor symmetry breaking, we introduce
parameter $R$ which describes
\begin{equation}
R\equiv  \langle (q\bar{s})_{V_1}(s\bar{q})_{V_2} |\hat{H}
|\chi_{c1}\rangle /\langle (q\bar{q})_{V_1}(q\bar{q})_{V_2} |\hat{H}
|\chi_{c1}\rangle \ ,
\end{equation}
and
\begin{equation}
R^2\equiv  \langle (s\bar{s})_{V_1}(s\bar{s})_{V_2} |\hat{H}
|\chi_{c1}\rangle / \langle (q\bar{q})_{V_1}(q\bar{q})_{V_2}
|\hat{H} |\chi_{c1}\rangle \ ,
\end{equation}
where the exchange of $V_1$ and $V_2$ is implicated, and the SU(3)
flavor symmetry is recognized by $R=1$. This parameter can be
related to the ratio of the $\pi$ and $K$ meson decay constant, i.e.
\begin{equation}
R\simeq f_\pi/f_K \ ,
\end{equation}
which gives $R\simeq 0.838$~\cite{Amsler:2008zzb}.

A commonly adopted vertex form factor is also applied,
\begin{equation}
{\cal {F}}^2({\bf p}_1) \equiv |{\bf p}_1|^{2L}\exp ({-{\bf
p}_1^2/{8\beta^2}}),
\end{equation}
where ${\bf p}_1$ is the three-vector momentum of the final-state
meson $V_1$ in the rest frame of $\chi_{c1}$, and $L$ is the
relative orbital angular momentum between $V_1$ and $V_2$. As
discussed earlier, $L=2$ is required by parity conservation and
Lorentz invariance. We adopt $\beta = 0.5 \ \mbox {GeV}$, which is
the same as in
Refs.~\cite{Close:2005vf,Zhao:2005im,Zhao:2007ze,Zhao:2008eg,Li:2007ky}.
The partial decay widths are given as follows:
\begin{eqnarray}
\Gamma_{\rho^+\pi^-}&= & \Gamma_{\rho^-\pi^+}=\frac{|{\bf
p}_1|}{24\pi M_{\chi_{c1}}^2} g_0^2 {\cal {F}}^2({\bf p}_1) \ ,
\nonumber\\
\Gamma_{K^{*0}\bar{K}^{*0}}&=&\Gamma_{\bar{K}^{*0}K^{*0}}=\Gamma_{K^{*+}K^{*-}}=\Gamma_{K^{*-}K^{*+}}=\frac{|{\bf
p}_1|}{24\pi M_{\chi_{c1}}^2} g_0^2 R^2 {\cal {F}}^2({\bf p}_1) \ ,
\nonumber\\
\Gamma_{\omega\omega}&=& \frac{|{\bf p}_1|}{24\pi M_{\chi_{c1}}^2}
g_0^2 {\cal {F}}^2({\bf p}_1) \ ,
\nonumber\\
\Gamma_{\phi\phi}&=& \frac{|{\bf p}_1|}{24\pi M_{\chi_{c1}}^2} g_0^2
R^4 {\cal {F}}^2({\bf p}_1) \ .
\end{eqnarray}

\end{document}